# Stability analysis of an oscillating tip–cantilever system in NC-AFM

G. Couturier, L. Nony, R. Boisgard, J.-P. Aimé*

*CPMOH, UMR5798 CNRS, Université Bordeaux I, 351, cours de la Libération, 33405 Talence Cedex, France*

### Abstract

This paper is a theoretical and a numerical investigation of the stability of a tip–cantilever system used in noncontact atomic force microscopy (NC-AFM) when it oscillates close to a surface. No additional dissipative force is considered. The theoretical approach is based on a variational method exploiting a coarse grained operation that gives the temporal dependence of the nonlinear-coupled equations of motion in amplitude and phase of the oscillator. Stability criterions for the resonance peak are deduced and predict a stable behavior of the oscillator in the vicinity of the resonance. The numerical approach is based on the results obtained with a virtual NC-AFM developed by our group. The effect of the size of the stable domain in phase is investigated. These results are in particularly good agreement with the theoretical predictions. Also they show the influence of the phase shifter in the feedback loop and the way it can affect the damping signal.

## 1. Introduction

In the recent years, the use of the noncontact atomic force microscopy (NC-AFM) mode has shown that contrasts at the atomic scale could be achieved on semiconductors and insulators surfaces [1–5]. Experimental and theoretical features dedicated to the description of this dynamic mode have been widely discussed in the previous papers [6–12]. In particular, it was shown that the high sensitivity of the oscillating tip–cantilever system (OTCS) was based on the value of the quality factor and on its nonlinear dynamics in the vicinity of the surface [10,13]. Current works of the authors focus on the origin of the increase of the damping signal when the tip comes close to the surface, in the range of a few angstroms. Some initial claims suggested that the origin of this apparent increase could be due to an hysteretic behavior of the OTCS [10,14]. These interpretations rise the question of the stability of the OTCS at the proximity of the surface. Recently, few authors have performed analytical and numerical works taking into account the time dependence of the equations of motion of the OTCS and have deduced stability criterions [15–17].

This work completes the analytical and numerical approaches developed in Refs. [16,17], respectively. Its aim is to show that the nonlinear dynamics of the OTCS leads to various stable branches on the resonance peak that may help to understand the reason why the NC-AFM mode, while being so sensitive, keeps a stable behavior. The influence of the phase shifter in the feedback loop and the way it can affect the damping signal is also investigated.

*Corresponding author. Tel.: +33-5-56-84-89-56; fax: +33-5-56-84-69-70.
*E-mail address:* jpaime@cribx1.u-bordeaux.fr (J.-P. Aimé).

## 2. Theoretical approach of the NC-AFM

### 2.1. Theoretical frame

To describe the interaction between the tip and the surface, the attractive coupling force is assumed to derive from a sphere–plane interaction involving the disperse part of the Van der Waals potential [18]:

$$V_{\text{int}}[z(t)] = -\frac{HR}{6[D - z(t)]} \quad (1)$$

where $H$, $R$ and $D$ are the Hamaker constant, the tip's apex radius, and the distance between the surface and the equilibrium position of the OTCS, respectively.

The complete theoretical approach of the stability of the OTCS was developed elsewhere [16]; here are given the main results. We search a solution to the temporal evolution of the OTCS by using a variational solution based on the principle of least action:

$$\delta S = \delta \left[ \int_{t_a}^{t_b} \mathscr{L}(z, \dot{z}, t) \, dt \right] = 0 \quad (2)$$

$\mathscr{L}$ is the Lagrangian of the system and $z(t)$ the position of the tip with time:

$$\mathscr{L}(z, \dot{z}, t) = \frac{1}{2} m^* \dot{z}(t)^2$$
$$- \left[ \frac{1}{2} k_c z(t)^2 - z(t) \mathscr{F}_{\text{exc}} \cos(\omega t) + V_{\text{int}}[z(t)] \right]$$
$$- \frac{m^* \omega_0}{Q} z(t) \underline{\dot{z}}(t) \quad (3)$$

where $\omega_0$, $Q$, $m^*$ and $k_c = m^* \omega_0^2$ are the resonance pulsation, quality factor, effective mass and cantilever stiffness of the OTCS, respectively. $\mathscr{F}_{\text{exc}}$ and $\omega$ are the external drive force and drive pulsation, respectively. Due to the large quality factor, we assume that a typical temporal solution is of the form:

$$z(t) = A(t) \cos[\omega t + \varphi(t)] \quad (4)$$

$A(t)$ and $\varphi(t)$ are assumed to be slowly varying functions with time compared to the period $T = 2\pi/\omega$. The underlined variables of $\underline{\dot{z}}(t)$ in Eq. (4), e.g. $\underline{A}(t)$, $\underline{\varphi}(t)$, $\underline{\dot{A}}(t)$ and $\underline{\dot{\varphi}}(t)$, are calculated along the physical path, thus they are not varied into the calculations [19]. The equations of motion in amplitude and phase of the OTCS are obtained by considering the following coarse-grained operation. Let us assume a long duration $\Delta t = t_b - t_a$ with $\Delta t \gg T$ and calculate the action as a sum of small pieces of duration $T$:

$$S = \sum_n \int_{nT}^{(n+1)T} \mathscr{L}(z, \dot{z}, t) \, dt$$
$$= \sum_n \left( \frac{1}{T} \int_{nT}^{(n+1)T} \mathscr{L}(z, \dot{z}, t) \, dt \right) T = \sum_n \mathscr{L}_e T \quad (5)$$

where $\mathscr{L}_e$ is the mean Lagrangian during one period and appears as an effective Lagrangian for a large time scale compared to the period. Owing to the quasi-stationary behavior of the amplitude and the phase over the period, the effective Lagrangian is calculated by keeping them constant during the integration. The calculations give

$$\mathscr{L}_e(A, \dot{A}, \varphi, \dot{\varphi})$$
$$= \frac{m^*}{4} [\dot{A}^2 + A^2(\omega + \dot{\varphi}^2)] - \frac{k_c A^2}{4} + \frac{\mathscr{F}_{\text{exc}} A \cos(\varphi)}{2}$$
$$- \frac{1}{T} \int_0^T V_{\text{int}}[z(t)] \, dt - \frac{m^* \omega_0}{2Q} [A \underline{\dot{A}} \cos(\underline{\varphi} - \varphi)$$
$$- A \underline{A}(\omega + \underline{\dot{\varphi}}) \sin(\underline{\varphi} - \varphi)] \quad (6)$$

Note that the effective Lagrangian is now a function of the new generalized coordinates $A$, $\varphi$ and their associated generalized velocities $\dot{A}$, $\dot{\varphi}$. At this point, remembering that the period is small regardless to $\Delta t = t_b - t_a$ during which the total action is evaluated, the continuous expression of the action is:

$$S = \int_{t_a}^{t_b} \mathscr{L}_e(A, \dot{A}, \varphi, \dot{\varphi}) \, d\tau \quad (7)$$

where the measure $d\tau$ is such that $T \ll d\tau \ll \Delta t$.

Applying the principle of least action $\delta S = 0$ to the functional $\mathscr{L}_e$, we obtain the Euler–Lagrange equations:

$$\ddot{a} = [(u + \dot{\varphi})^2 - 1]a - \frac{\dot{a}}{Q} + \frac{\cos(\varphi)}{Q} + \frac{a\kappa_a}{3(d^2 - a^2)^{3/2}},$$
$$\ddot{\varphi} = -\left(\frac{2\dot{a}}{a} + \frac{1}{Q}\right)(u + \dot{\varphi}) - \frac{\sin(\varphi)}{aQ} \quad (8)$$

In the above equation, specific notations were used. $d = D/A_0$ is the reduced distance between the location of the surface and the equilibrium position of the OTCS normalized to the resonance amplitude $A_0 = Q\mathscr{F}_{\text{exc}}/k_c$. $a = A/A_0$ is the reduced amplitude, $u = \omega/\omega_0$ the reduced drive frequency normalized to the resonance frequency of the free OTCS and $\kappa_a = HR/k_c A_0^3$ is the dimensionless parameter that characterizes the strength of the interaction.

## 2.2. Resonance frequency shift

The equations of motion of the stationary solutions $a$ and $\varphi$ are obtained by setting $\dot{a} = \dot{\varphi} = 0$ and $\ddot{a} = \ddot{\varphi} = 0$ in Eq. (8) and lead to two-coupled equations of the sine and cosine of the phase of the OTCS previously calculated [20]:

$$\cos(\varphi) = Qa(1-u^2) - \frac{aQ\kappa_a}{3(d^2-a^2)^{3/2}},$$
$$\sin(\varphi) = -ua \qquad (9)$$

Solving the above equation gives the relationship between the frequency and the amplitude at a given distance $d$ [10]:

$$u_{\pm}(a) = \sqrt{\frac{1}{a^2} - \frac{1}{4Q^2}\left(1 \mp \sqrt{1 - 4Q^2\left(1 - \frac{1}{a^2} - \frac{\kappa_a}{3(d^2-a^2)^{3/2}}\right)}\right)^2} \qquad (10)$$

The signs plus and minus are deduced from the sign of $\cos(\varphi)$ and correspond to values of the phase ranging from 0 to $-\pi/2$ ($u_-$, $\cos(\varphi) > 0$) or from $-\pi/2$ to $-\pi$ ($u_+$, $\cos(\varphi) < 0$), in agreement with the sign convention of the phase in Eq. (4). From Eq. (10) is calculated the resonance peak at any reduced distance for a given strength of the sphere–surface interaction, and Eq. (9) gives the associated phase variations. The two branches define the distortion of the resonance peak as a function of $d$. $u_-$ gives the evolution of the resonance peak for frequency values below the resonance one and $u_+$ for frequency values above the resonance (see Fig. 1(a)).

Using Eq. (10), the resonance frequency shift as a function of the distance $d$ is obtained by setting $a = 1$. This former condition ensures the required condition for the NC-AFM mode. Thus, the normalized frequency shift $(v - v_0)/v_0$ is given by $u - 1$ [10]:

$$u_{\pm}(d) - 1 = \sqrt{1 - \frac{1}{4Q^2}\left(1 \mp \sqrt{1 + \frac{4}{3}\frac{Q^2\kappa_a}{(d^2-1)^{3/2}}}\right)^2} - 1 \qquad (11)$$

As $d$ decreases, the branches $u_+$ and $u_-$ become closer and closer in the vicinity of the resonance. Therefore question rises about the ability of the OTCS to remain on the same branch. Qualitatively, one may expect that around $a \cong 1$, the branch $u_-$ is unstable and $u_+$ is stable (see Fig. 1(a)). If this is true, any small fluctuation of the oscillation amplitude might produce a jump from one branch to the other one as discussed in Refs. [10,14]. Since the branch $u_-$ seems to be unstable, a jump to this branch should lead to an abrupt decrease of the amplitude, which in turn might produce an apparent abrupt increase of the damping signal as a consequence of the nonlinear behavior.

Because such a jump is never observed, it becomes useful to determine more accurately the stability of the two branches.

## 2.3. Stability criterions

The stability of the branches $u_{\pm}$ is deduced by linearizing the equations of motion of the OTCS (see Eq. (8)) around the stationary solution (now identified by the index "s") and using classical considerations of the linear theory. A similar approach leading to equivalent results was developed by Gauthier et al. [15]. The stability criterions can be expressed from the derivatives $da_s/du_{\pm}$ of the branches and reduced to the simple expression [16]:

$$\frac{da_s}{du} > 0 \quad \text{and} \quad \cos(\varphi_s) > \frac{a_s}{(2Q)} \quad \text{or} \qquad (12a)$$

$$\frac{da_s}{du} < 0 \quad \text{and} \quad \cos(\varphi_s) < \frac{a_s}{(2Q)} \qquad (12b)$$

Figs. 1(a) and (b) show the distortion of the resonance peak and associated phase curve, respectively. Figs. 2(a) and (b) are zooms on the region $\alpha$ of Figs. 1(a) and (b), respectively.

For the branch $u_+$, $da_s/du_+$ being always negative and the associated value of the phase being always defined beyond $-\pi/2$ (see Section 2.2), the criterion (12b) implies that $u_+$ is always stable, whatever the value of $a_s$.

For $u_-$, the sign of the derivative changes twice. For this branch, the phase is always defined above $-\pi/2$. Therefore on the lower part of the branch

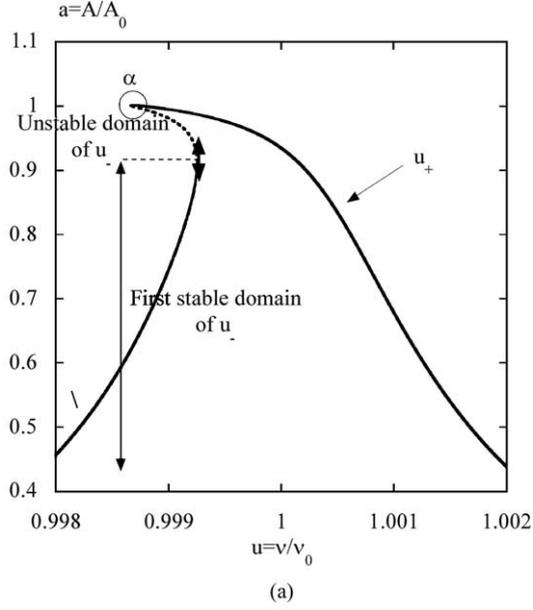

(a)

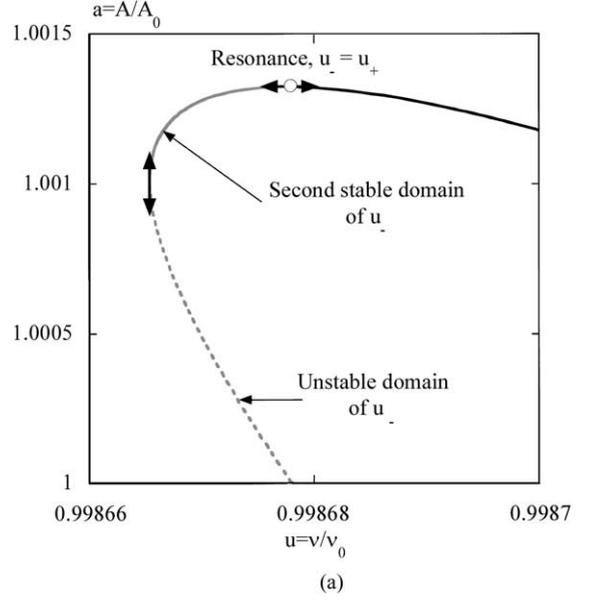

(a)

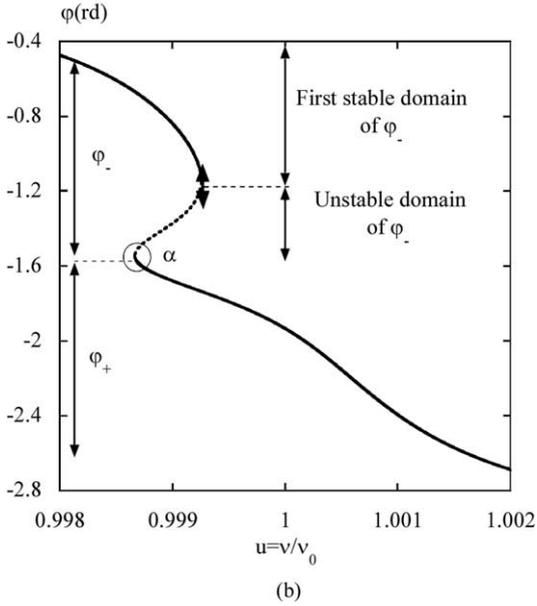

(b)

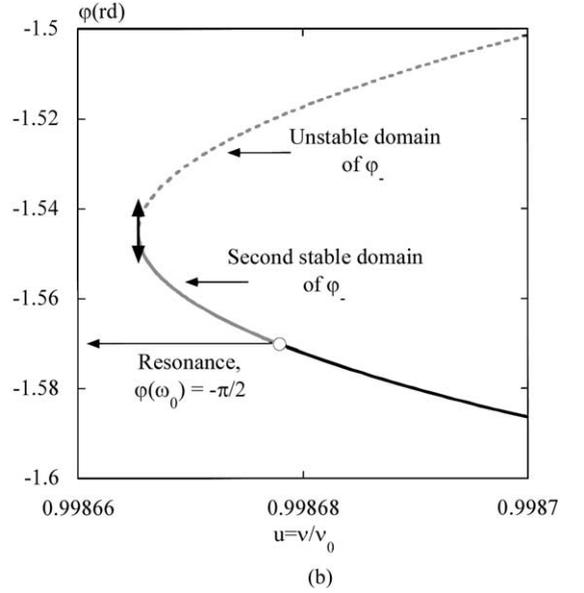

(b)

Fig. 1. (a) Distortion of the resonance peak computed from Eq. (10). The numerical parameters are $d = 1.05$, $A_0 = 10$ {nm}, $Q = 500$ and $\kappa_a = 2.5 \times 10^{-4}$. The stability criterions foresee that $u_+$ is always stable, whereas $u_-$ exhibits two stable domains (continuous lines, see also Fig. 2) and one unstable (dashed lines). The domains are separated by the spots where the derivative $da/du_-$ diverges (see text). (b) Distortion of the phase curve computed from Eq. (9) associated to the resonance peak. As a consequence of the stability of $u_+$, $\varphi_+$ is always stable whereas $\varphi_-$ exhibits two stable domains and one unstable.

Fig. 2. (a) Zoom in the region α of the resonance peak given in Fig. 1. As $da/du_-$ diverges again, this defines a new domain of $u_-$ which is predicted to be stable (see text). The resonance is located where $u_+ = u_-$. (b) Zoom in the region α of the phase curve. The resonance is located at $-\pi/2$ where $\varphi_+ = \varphi_-$ and belongs to a stable domain.

(small $a$), $da_s/du_- > 0$ and the criterion (12a) indicates that the branch is locally stable. When $da_s/du_-$ becomes negative (see Fig. 1(a)), because $\varphi_s > -\pi/2$ (see Fig. 1(b)), the criterion (12a) is no more filled.

As a consequence, $u_-$ is locally unstable and the instability is precisely located where the infinite tangent appears. On the upper part of the resonance peak, the curvature of $u_-$ changes again and $da_s/du_- > 0$ (see Fig. 2(a)), implying that it is again a locally stable domain. Thus the branches $u_-$ and $\varphi_-$ exhibit two stable domains and one unstable.

Note also that the resonance condition is deduced from $da_s/du = 0$ which implies $\cos(\varphi_s) = a_s/2Q$, or equivalently $u_- = u_+$, or again $\varphi_- = \varphi_+$. This equality is the usual resonance condition of a free harmonic oscillator. If $a_s = 1$, e.g. without any coupling, the resonance phase is therefore $\varphi_s = \arccos[1/2Q]$. For the OTCS we used, $Q \simeq 500$, and so $\varphi_s \cong -\pi/2$. But taking into account the fact that the coupling only slightly modifies the value of the resonance amplitude, $a_s \simeq 1.0013$ (see Fig. 2(a)), we still obtain $\varphi_s \cong -\pi/2$ so that we can consider that the nonlinear resonance is always given by the relationship $\varphi_s = -\pi/2$.

The theoretical approach foresees that $u_+$ is always stable but that also a small domain of $u_-$ around the resonance value remains stable. If the resonance value would have been located at the point where $da_s/du_-$ is infinite, an infinitely small fluctuation would have been able to generate an abrupt increase of the damping signal as discussed previously and suggested in Ref. [10], or more recently in Ref. [14]. Experimentally, an electronic feedback loop keeps constant the amplitude of the OTCS so that its phase is located around $-\pi/2$ (see the below section). As a consequence, question rises about the size of the stable domain in phase around $-\pi/2$. If any fluctuation around $-\pi/2$ makes the phase going beyond the stable domain, the OTCS behavior becomes unstable. For $Q = 500$, the size of the stable domain is of about $2.6 \times 10^{-2}$ rd (see Fig. 2(b)) whereas it is reduced to $2.6 \times 10^{-3}$ rd for $Q = 5000$ (data not shown). Thus, if the electronic loop is able to control the phase locking with a better accuracy than $2.6 \times 10^{-3}$ rd, the OTCS will be locked on a stable domain.

## 3. Virtual NC-AFM results

In a recent paper, we have described a virtual NC-AFM machine built using the *Matlab* language and the *Simulink* toolbox [21]. This machine is very similar to our own experimental hybrid machine built with *Digital Instruments* [22] and *Omicron* [23] blocks. The virtual machine has been extensively used to study the frequency shift and the damping signal in the approach-retract mode.

Here we want to use the virtual machine to investigate the stability of the OTCS by looking accurately at its phase variations within the electronic feedback loop that maintains constant the amplitude of the oscillations.

### 3.1. The phase shifter of the feedback loop

In Fig. 3 is drawn a simplified schematic diagram of the feedback loop of our NC-AFM (for more details, see Ref. [21]). Usually, the phase $\phi(\omega)$ of the phase shifter transfer function is adjusted to $-3\pi/2$ so that the loop oscillates at $\nu_0$ which is the free resonance frequency of the cantilever, corresponding to a tip–surface distance $D \to \infty$. The oscillations of the loop are ruled by the relation:

$$\phi(\omega) + \varphi(\omega) = 0 \pm 2n\pi \qquad (13)$$

where $n$ is an integer and $\varphi(\omega)$ the phase difference between the oscillations and the excitation of the cantilever. If the set point is fixed to the resonance frequency, then $\varphi(\omega_0) = -\pi/2$. The phase adjustment in the *Omicron* electronics is obtained by changing the bias of varicap diodes [24]. The phase shifter transfer

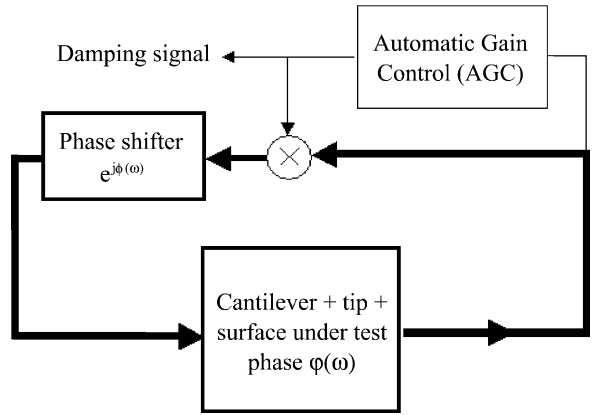

Fig. 3. Schematic diagram of the feedback loop used in the virtual NC-AFM which is very similar to the one of the experimental machine.

function in terms of the $p$ Laplace variable can be written as

$$H(p) = \left(\frac{1-\tau p}{1+\tau p}\right)^2 \quad (14)$$

The time constant $\tau$ is experimentally adjusted such that, at the resonance

$$\phi(\omega_0) = -4\arctan(\tau\omega_0) = -\frac{3\pi}{2} \quad (15)$$

When the tip–surface distance $D$ is reduced, due to the coupling, $\omega_0$ decreases. As a consequence, $\phi(\omega_0)$ and $\varphi(\omega_0)$ are no more equal to $-3\pi/2$ and $-\pi/2$, respectively. According to Eq. (13), the variation of $\varphi(\omega)$ is governed by the one of $\phi(\omega)$. Assuming a small variation around the resonance frequency $\Delta\omega = \omega - \omega_0$, one gets

$$\phi(\omega) \simeq \frac{-3\pi}{2} - \frac{4\tau}{1+(\tau\omega_0)^2}\Delta\omega \quad (16)$$

As $D$ decreases, $\Delta\omega$ is negative. Therefore $\phi(\omega)$ becomes larger than $-3\pi/2$ and $\varphi(\omega)$ smaller than $-\pi/2$. The decrease of $\varphi(\omega)$, $\varphi(\omega) \lesssim -\pi/2$, means that the phase of the OTCS follows the phase branch associated to $u_+$, $\varphi_+$, which is always stable (see Fig. 2(b)). Thus the loop is always stable. Moreover, the hypothesis implying that $\varphi(\omega)$ keeps a value close to $-\pi/2$ is a very good assumption. To prove that, let us consider for instance $v_0 = 150$ kHz, therefore $\tau = 2.56 \times 10^{-6}$ s (see Eq. (15)). Assuming now a large frequency shift, $\Delta v = -200$ Hz, we get $\Delta\phi(\omega) = 1.9 \times 10^{-3}$ rd and therefore $\Delta\varphi(\omega) = -1.9 \times 10^{-3}$ rd. Thus the typical phase variations of the OTCS around the nonlinear resonance are less than $2 \times 10^{-3}$ rd. This implies that the machine properly follows the nonlinear resonance, even when large frequency shifts are considered.

The curve [a] in Figs. 4 and 5 shows the phase $\varphi(\omega)$ and the damping signal vs. the distance $D$, respectively. As expected, the variations are very weak.

### 3.2. "Controlled" damping variations

To observe the phase instability predicted by the theoretical calculations, the phase shifter transfer function should have been in the form $d\phi(\omega)/d\omega > 0$ around $\omega_0$. A possible expression of such a transfer function would be $H(p) = ((1+\tau p)/(1-\tau p))^2$. Experimentally, this form is not feasible and even if it were, the loop would become unstable and therefore no stationary state could be reached. The reason is that the inverse Laplace transform of $1/(1-\tau p)$ varies as $e^{t/\tau}$ which diverges as $t \to \infty$.

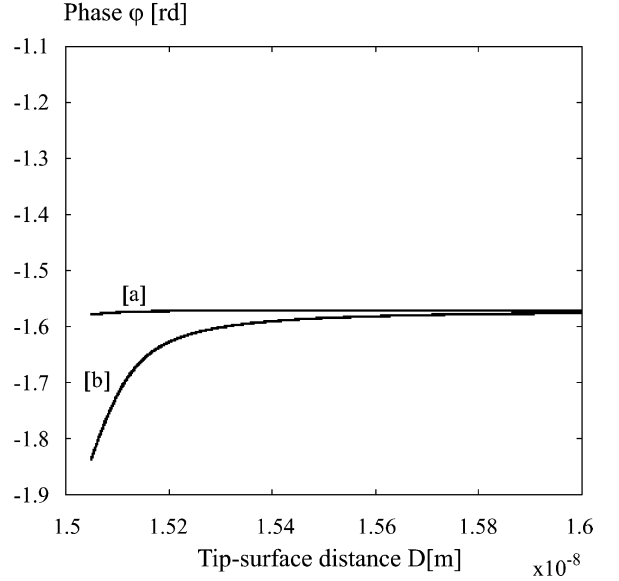

Fig. 4. Variations of the phase of the OTCS $\varphi(\omega)$ within the feedback loop vs. the distance $D$ computed from the virtual NC-AFM. The numerical parameters are: resonance amplitude $A_0 = 15$ nm, spring constant $k_c = 40$ N m$^{-1}$, quality factor $Q = 5000$, tip's radius $R = 10$ nm and Hamaker constant $H = 2 \times 10^{-19}$ J. Curve [a]: the phase $\phi(\omega)$ of the transfer function $H(p)$ of the phase shifter is the one given by Eq. (15), e.g. is similar to the experimental machine. As $D$ decreases, $\varphi(\omega)$ becomes weakly smaller than $-\pi/2$ (less than $2 \times 10^{-3}$ rd, see text), therefore follows the stable branch $\varphi_+$. The machine follows accurately the set point, which is always stable, even when the tip is in the very close vicinity of the surface. The associated damping variation nearly not varies (see Fig. 5). Curve [b]: $\phi(\omega)$ is the phase of $H(p)$ whose expression is given by Eq. (17). Around $-\pi/2$, the slope $d\phi(\omega)/d\omega$ is larger than in curve [a] so that $\varphi(\omega)$ decreases more quickly. As a consequence, the damping signal increases.

Nevertheless, it is possible to exploit the phase instability to investigate more accurately the damping signal variations with the virtual machine. In the virtual machine, it is possible to implement a phase shifter with a slope $d\phi(\omega)/d\omega$ larger than the one built by *Omicron*. As an example, we have retained a transfer function which is easy to do with electronic components

$$H(p) = \left(\frac{p^2 - (\omega_1/Q_1)p + \omega_1^2}{p^2 + (\omega_1/Q_1)p + \omega_1^2}\right) \quad (17)$$

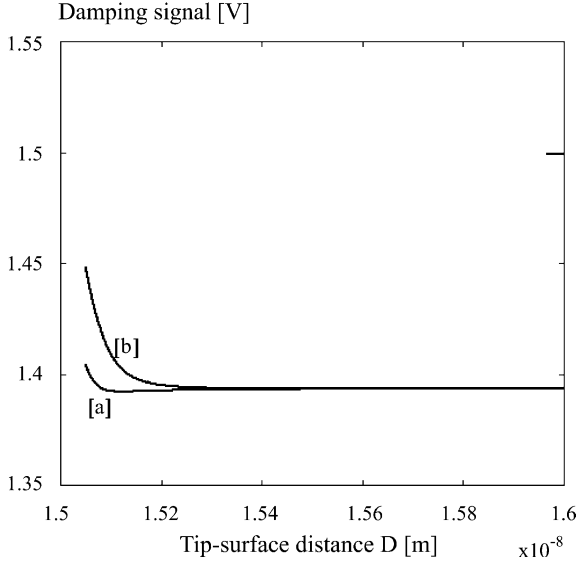

Fig. 5. Variations of the damping signal vs. the distance $D$ correlated to the phase variations given in Fig. 4. Curve [a]: no damping variation is observed if the phase of the virtual NC-AFM phase shifter is similar to the one of the experimental machine. In the very close vicinity of the surface, as $\varphi(\omega_0) - \pi/2$, e.g. the amplitude of the oscillations slightly decreases, a weak increase is observed. Curve [b]: as $\varphi(\omega_0)$ varies more quickly than in curve [a] due to the different expression of $H(p)$, a larger increase of the damping is obtained.

The parameters $\omega_1$ and $Q_1$ may be adjusted to obtain for instance $\phi(\omega_0) = -3\pi/2$. The phase of the transfer function is then:

$$\phi(\omega) = -2\arctan\left(\frac{\omega_1\omega}{Q_1(\omega_1^2 - \omega^2)}\right) \quad (18)$$

For a small frequency shift, $d\phi(\omega)/d\omega|_{\omega_0} \simeq -4Q_1/\omega_0$. Keeping the same values than previous $\nu_0 = 150$ kHz and $\Delta\nu = -200$ Hz and assuming $Q_1 = 50$, we now obtain a change $\Delta\phi(\omega)$ of about 0.26 rd. Consequently, $\Delta\varphi(\omega)$ becomes larger (see the curve [b] in Fig. 4) and we now observe an increase of the damping signal as shown in Fig. 5, curve [b].

The previous examples are pedagogical cases. The latter one considered an arbitrary large value of the phase slope of the phase shifter. An ideal phase shifter should maintain the phase $\phi(\omega_0)$ at $-3\pi/2$ so that the frequency of the loop remains equal to the resonance frequency of the cantilever. Practically, this is not possible. However, it is clear that the solution retained by *Omicron* is very close to the ideal case $\phi(\omega_0) = -3\pi/2$ because $d\phi(\omega)/d\omega$ is very weak.

## 4. Conclusion

A variational method based on a coarse-grained operation has been used to investigate in details the stability of an oscillating tip–cantilever system near a surface. The tip–surface interaction is described by Van der Waals potential. Results show that the resonance peak of the oscillator can be described from two branches. The first one, named $u_+$, corresponds to frequencies larger than the resonance. Stability criterions deduced foresee that it is always stable. The second one, $u_-$, may be decomposed into three domains: two are stable and one is unstable. The second stable domain of $u_-$ is small and is defined at the upper extremity of the resonance peak. The phase at the resonance $\varphi(\omega_0) = -\pi/2$ is at the overlap of the $u_+$ and of this former second stable domain of $u_-$, thus the set point $\varphi(\omega_0) = -\pi/2$ belongs to a stable zone.

This result is of great importance to understand the stability in NC-AFM. In this technique, the phase of the cantilever is adjusted to $-\pi/2$ within an electronic feedback loop as the tip–surface distance is large enough to make the tip nonsensitive to the interaction. In the approach mode, the frequency of the loop decreases, consequently the phase becomes smaller than $-\pi/2$ because the phase slope $d\phi(\omega)/d\omega$ of the phase shifter transfer function is always negative. Thus the oscillator always "slides" along $u_+$ and the system is unconditionally stable. This is what usually observed experimentally and confirmed by the results of the virtual NC-AFM we have built. Because the slope $d\phi(\omega)/d\omega$ and the frequency shift are very weak, we may consider that the phase $\varphi(\omega_0)$ of the oscillator is always very close to $-\pi/2$, typical variations being less than $2 \times 10^{-3}$ rd. Consequently, the damping signal keeps constant if no dissipative force is introduced in the tip–surface interaction.